\documentclass{aa}  
\usepackage{graphicx}
\begin{document}
\title{The Mass of the Planet-hosting Giant Star 
$\beta$ Geminorum Determined from its p-mode Oscillation Spectrum}


   \author{A. P. Hatzes\inst{1}
\and
	M. Zechmeister\inst{2}
          \and
	J. Matthews\inst{3}
	\and
	R. Kuschnig\inst{4}
	\and
\\
	G.A.H. Walker\inst{5}
        \and
   M. D\"ollinger\inst{1}
        \and
  D.B. Guenther\inst{6}
        \and
A.F.J. Moffat\inst{7}
        \and
S.M. Rucinski\inst{8}
        \and
D. Sasselov\inst{9}
         }

   \offprints{
    Artie Hatzes, \email{artie@tls-tautenburg.de}\\$*$~
    Based on observations obtained at the
    2m Alfred Jensch Telescope at the Th\"uringer
    Landessternwarte Tautenburg and 
data from the MOST satellite, a Canadian Space Agency mission, jointly operated by Dynacon Inc., 
the University of Toronto Institute 
of Aerospace Studies and the University of British Columbia with the 
assistance of the University of Vienna.
  \\}

     \institute{Th\"uringer Landessternwarte Tautenburg,
                Sternwarte 5, D-07778 Tautenburg, Germany
\and
Georg-August-Universit\"at G\"ottingen, Friedrich-Hund-Platz 1, 37077
\and
Department of Physics, University of British Columbia, 6224 Agricultural Road, Vancouver, BC V6T 1Z1, Canada
\and
Institut f\"ur Astronomie, Universit\"at Wien 
T\"urkenschanzstrasse 17, A-1180 Vienna, Austria
\and
1234 Hewlett Place, Victoria, BC V8S 4P7, Canada
\and 
Institute for Computational Astrophysics, Department of Astronomy
and Physics, Saint Mary's University, Halifax, NS B3H 3C3, Canada
\and
Observatoire Astronomque du Mont M\'egantic, D\'epartment de
Physique, Universit\'e de Montr\'eal C. P. 6128, Succursale :
Centre-Ville, Montr\'eal, QC H3C 3J7, Canada
\and
Department of Astronomy and Astrophysics, University of Toronto,
Toronto, ON M5S 3H4, Canada
\and
Harvard-Smithsonian Center for Astrophysics, 60 Garden Street,
Cambridge, MA 102138
}

   \date{Received; accepted}

 
  \abstract
{}
{Our aim is to use precise radial velocity measurements and
photometric data to derive the frequency spacing of the p-mode oscillation spectrum of the
planet-hosting star $\beta$ Gem. This  spacing along with the
interferometric radius for this star can then be used to derive an accurate  
stellar mass.
   }
{We use a  long time
series of over 60 hours of
precise stellar radial velocity measurements
of $\beta$ Gem taken with an iodine absorption cell at the echelle spectrograph
mounted on the 2m Alfred Jensch Telescope. 
We also present complementary photometric data for this star
taken with the MOST microsatellite spanning 3.6 d. A Fourier 
analysis is used to derive the frequencies that are present in each data set. 
}
{
The Fourier analysis of the radial velocity  data reveals the presence of
up to 17 significant pulsation modes in the frequency interval 10--250
$\mu\mathrm{Hz}$. Most of these fall on a grid of equally-spaced
frequencies having a separation of 7.14 $\pm$ 0.12 $\mu\mathrm{Hz}$.
An analysis of 3.6 days of high precision photometry
taken with the MOST space telescopes  shows the presence of up to
16 modes, six of which are consistent with modes found in the spectral 
(radial velocity) data.
This frequency  spacing is
consistent with  high overtone radial pulsations; however, until
the pulsation modes are identified we cannot be sure if some of these
are nonradial modes or even mixed modes.  The radial velocity
 frequency spacing
along with angular diameter measurements of $\beta$ Gem via
interferometry results in a stellar mass of M = 1.91 $\pm$ 0.09 M$_\odot$.
This value confirms the intermediate mass of the star
determined using stellar  evolutionary tracks.
}
{$\beta$ Gem is confirmed to be an intermediate mass star.
Stellar pulsations in giant stars along with interferometric
radius measurements can provide accurate determinations of the stellar mass
of planet hosting giant stars. These can also be used to calibrate
stellar evolutionary tracks.
}

\titlerunning{The p-mode oscillation spectrum of $\beta$ Gem}
\maketitle
%

\section{Introduction}

There are several programs underway that use precise stellar radial velocity
(RV) measurements to search for planets around evolved giant stars
(e.g. Setiawan et al. 2003; Sato et al.  2003, Johnson et al. 2007; Niedzielski et al. 2009;
Han et al. 2010). The reason is that the progenitors
of these giant stars are A-F dwarfs -- stars that have high
effective temperatures and rapid rotation rates that  makes it almost impossible
to detect the subtle reflex motion from any planetary companion with
RV measurements. Although giant stars offer us a means of detecting planetary
companions around stars more massive than the sun, there is a drawback 
in the determination of an accurate stellar mass. Unlike  for main sequence stars
that have a tight and reasonably well calibrated relationship between stellar
mass and effective temperature, the evolutionary tracks of giants stars covering
a broad range of stellar masses all converge to the same region of the H-R
diagram. One must rely on evolutionary tracks, which are model dependent and require
accurate stellar parameters (luminosity, effective temperature, abundance, etc.), to determine
the stellar mass. Accurate stellar masses of planet hosting giant stars are 
needed if we are to understand the process of planet formation as a function of stellar
mass.

\begin{figure*}
\resizebox{\hsize}{!}{\includegraphics{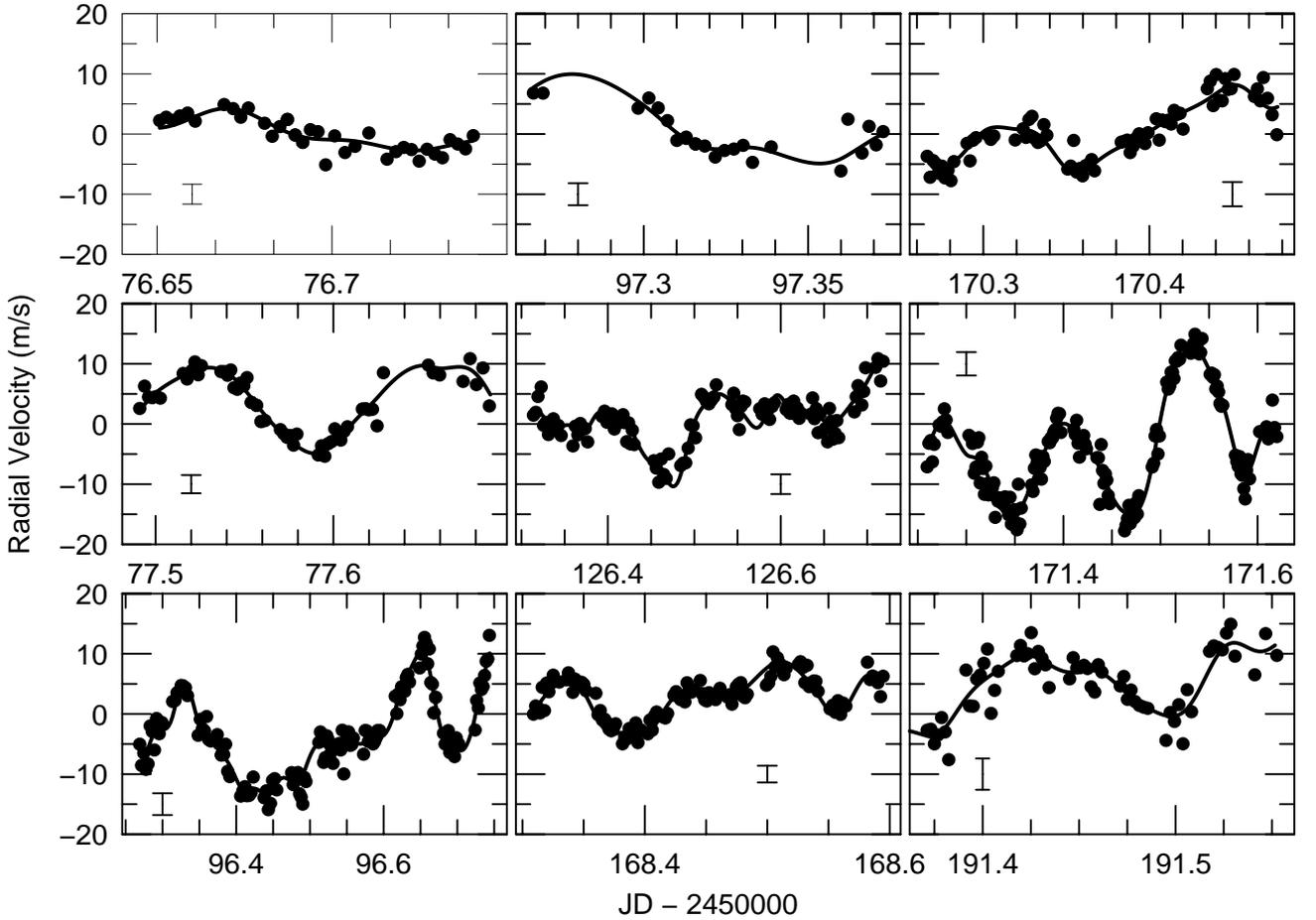}}
\caption{Radial velocity measurements for $\beta$ Gem on 9 different nights.
The error bar in each panel shows the typical error for each night.
The solid line represents the fit to the data using the frequencies and
amplitudes in Table 2. For clarity we used variable limits
for the ordinate.
}
\label{timeseries}
\end{figure*}

The bright star $\beta$ Gem is a planet-hosting K giant star 
(Hatzes \& Cochran 1993; Hatzes et al. 2006; Reffert et al. 2006)
that was recently shown to exhibit stellar oscillations. 
Hatzes \& Zechmeister (2007; hereafter HZ07) using 20 hr of precise stellar radial velocity
measurements demonstrated that this star shows up to 6 oscillation modes
in the frequency range $\nu$ = 10 -- 150 $\mu\mathrm{Hz}$.
The most dominant mode was at $\nu$ = 
89 $\mu\mathrm{Hz}$ and had an amplitude of 5 m\,s$^{-1}$. The pulsation frequencies
and amplitudes were consistent with those expected for solar-like p-mode
oscillations for a giant star with radius $R$ = 8.8 $R_\odot$ and mass, 
$M$ $\approx$  2 $M_\odot$.  The former was determined from interferometric
measurements (Nordgren et al. 2001).

The mass estimate of $\beta$ Gem by HZ07 was 
based on the frequency of maximum power, $\nu_{max}$, and
the scaling relations of Kjeldsen \& Bedding (1995). This is not as accurate
for two reasons. First, using a short time span of data
(3 nights in the case of HZ07) we cannot be sure that the dominant mode we 
detect accurately represents $\nu_{max}$. Likewise fitting a Gaussian
to the envelope of observed excess power may have a large error. Second,
to derive the mass from $\nu_{max}$ one requires a knowledge of the effective
stellar temperature which is not well constrained. On the other hand,
the stellar mass determined from the frequency spacing depends only on the
stellar radius (measured with interferometry in this case) and is thus independent
of other more poorly determined stellar parameters like temperature or surface
gravity.  If one knows
the stellar radius, one can thus derive the stellar mass. 
Just as important,
determining the precise frequencies of as many detected modes as possible
and their separation
are more valuable for asteroseismic modeling than just determining $\nu_{max}$
Unfortunately, the data set of 
HZ07 was of insufficient length to do this.

In this paper we use a much larger data set with over 60 hr of coverage, including
the measurements presented in HZ07, 
to derive the oscillation spectrum and thus
the frequency spacing of $\beta$ Gem. We also present photometric
observations taken with the MOST space telescope.

\section{Observational Data}
\subsection{Spectral Data}

	Radial velocity measurements were made using the coude echelle 
spectrograph of the 2m Alfred Jensch Telescope of the Th\"uringer 
Landessternwarte Tautenburg (Thuringia State Observatory). Observations
were made on 10 nights over the time span 24 December 2006 to 
18 April 2007, or a difference of 116 days from first to last
observation. A total of over 60 hrs of  observations were
made resulting in over 762 measurements.  Precise stellar radial velocities were determined using 
an iodine absorption cell placed in the optical path of the telescope.
A further description of the instrumental setup and RV analysis method can be
found in Hatzes et al.  (2005). Table 1  lists the journal of observations.

\begin{table}[h]
\begin{center}
\begin{tabular}{cccc}
Start        & Time Coverage  & $N_\mathrm{Obs}$   &   $\sigma$ \\
(Julian Day) & (hours)        &              &   (m\,s$^{-1}$) \\
\hline\hline
2450076.650 & 2.16 &  38 & 1.65 \\
2450077.491 & 4.70 &  58 & 1.51 \\
2450096.269 &11.38 & 135 & 1.81 \\
2450097.266 & 2.54 &  24 & 1.86 \\
2450126.314 & 9.70 & 104 & 1.64 \\
2450168.308 & 6.88 & 115 & 1.39 \\
2450170.266 & 5.04 &  74 & 2.00\\
2450171.259 & 8.64 & 152 & 1.94\\
2450191.371 & 4.34 &  62 & 2.64\\
2450192.277 & 6.52 &  39 & 3.13\\
\hline
\end{tabular}
\caption{The journal of observations including Julian Date as the
start of the night, time coverage per night, number of observations each
night, and the rms scatter of the data about the multi-component frequency fit
(see text).}
\end{center}
\label{journal}
\end{table}

\begin{figure}
\resizebox{\hsize}{!}{\includegraphics{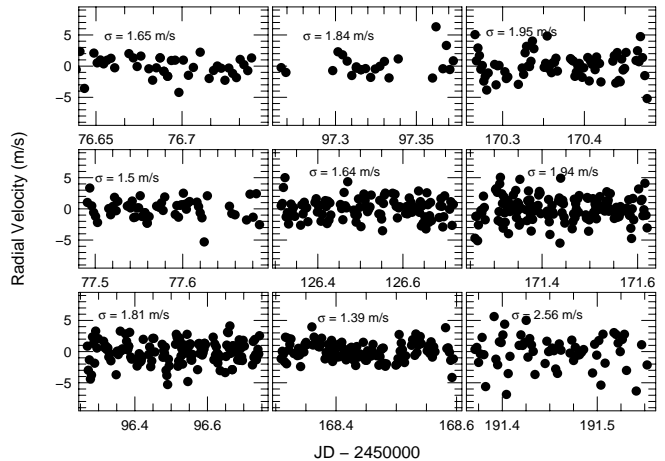}}
\caption{Residual radial velocities after subtracting the fit 
using the frequencies and
amplitudes in Table~\ref{rvmodes}. In each panel the rms
scatter for that night is given.
}
\label{resids}
\end{figure}

The planetary companion to $\beta$ Gem causes a reflex motion of 
39 m\,s$^{-1}$ with a period of 590 days. This orbital motion was removed
and the subsequent analysis performed on the residual RV data.
Figure~\ref{timeseries} shows the RV measurements for 9 of the
nights after removal of the orbital motion. Figure~\ref{resids} 
shows the residual RVs after subtracting the fit discussed in more detail
below. In order
to make a more compact figure for easy viewing we do not show the
last night which was also included in our analysis. Although the time
span of the observations (time between first and last data point) for 
this night is large
compared to the other nights, there were large gaps and only 39 measurements
were made due to poor observing conditions. Also, because of these conditions
the RV precision was about a factor of 2 worse than the other nights.
If we exclude these measurements the following results are not significantly altered.

\subsection{Photometric Data}
Photometric measurements were taken using the {\it Microvariability
and Oscillations of Stars Microsatellite} (MOST) space mission
as part of a secondary observational program. 
MOST is 15/17.3 cm Rumak-Maksutov telescope with a CCD photometer
in a circular Sun-synchronous polar
orbit with a period of 101.41 minutes (Walker et al. 2003).
MOST observed $\beta$ Gem on 5 consecutive nights spanning 13 -- 17 February 2007 
(JD = 2,454,145.633 and 2,454,149.26).  Individual exposures were set to 1.5 secs
and 14 images were stacked on board before downloading the data. Observations
of $\beta$ Gem alternated with those of a primary program of  
MOST. $\beta$ Gem was continuously observed for 40 min followed by a gap
of approximately 1 hour for a duty cycle of about 40\%. 
 A total of 4440 photometric measurements were made of
$\beta$ Gem

\begin{figure}
\resizebox{\hsize}{!}{\includegraphics{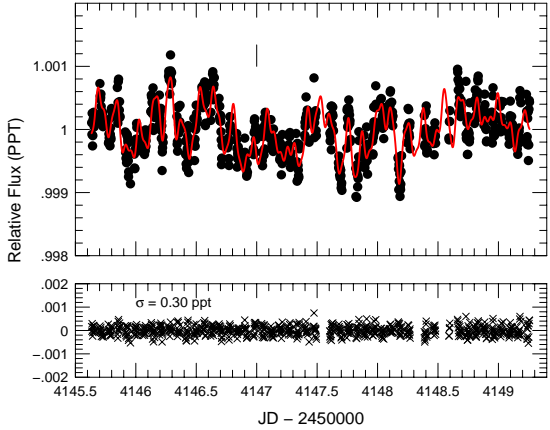}}
\caption{(Top) Photometry of $\beta$ Gem taken with the MOST micro-satellite.
The solid line represents the fit to the data using the frequencies and
amplitudes in Table~4. (Bottom) Residuals after subtracting the fit. Units are in
parts per thousand (PPT).
}
\label{mostphot}
\end{figure}

Figure~\ref{mostphot} shows the time series of the
photometric measurements.
For clarity data points shown in the figure are average values of 4
measurements (50 secs). 

\begin{figure}
\resizebox{\hsize}{!}{\includegraphics{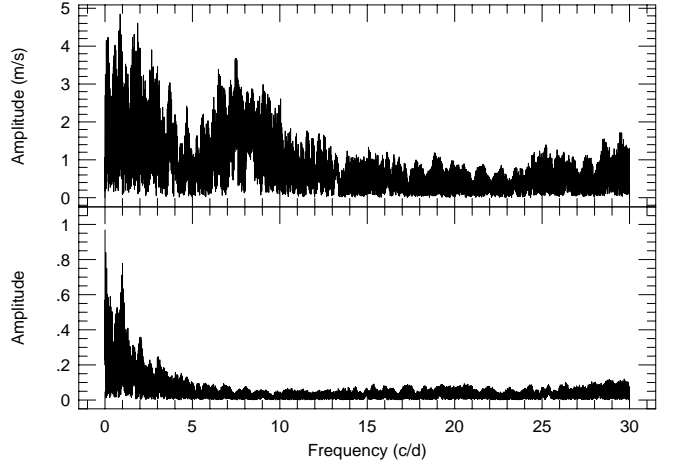}}
\caption{(Top) The amplitude spectrum for the RV data. (Bottom) the window
function.
}
\label{dft}
\end{figure}

\begin{figure}
\resizebox{\hsize}{!}{\includegraphics{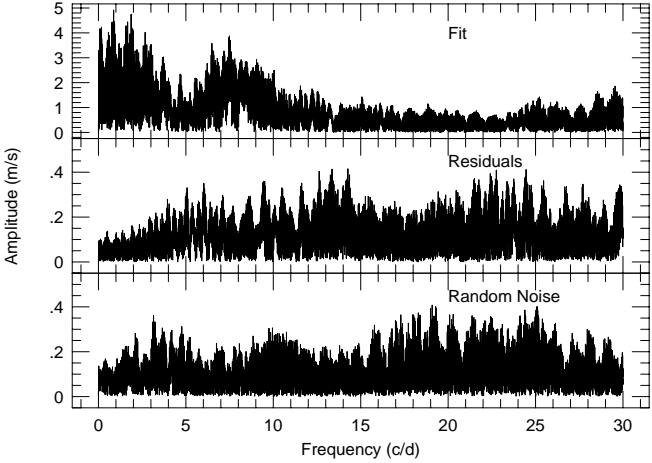}}
\caption{(Top) The amplitude spectrum of the 17-component fit to the
data that was sampled in the same manner as the data. Random noise
with $\sigma$  = 2 m\,s$^{-1}$ was also added to the synthetic data. 
(Middle) The amplitude spectrum of the residual RVs after subtracting the
fit to the data. (Bottom) The amplitude spectrum of random noise
with $\sigma$ = 2  m\,s$^{-1}$ that was sampled in the same manner as
the data. 
}
\label{final}
\end{figure}

\begin{figure}
\resizebox{\hsize}{!}{\includegraphics{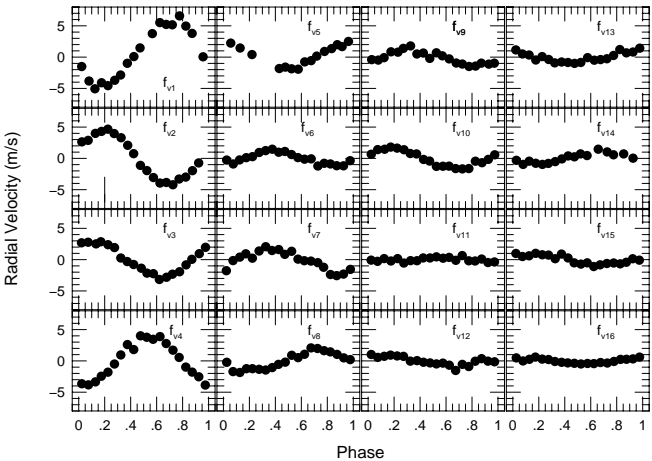}}
\caption{Phase diagrams showing the residual RV measurements phase
to the frequencies found in the Fourier analysis. For each panel the contribution
of all other frequencies were subtracted prior to phasing the data. Data were
phased in phase bins of 0.05. The gap in panel $f_{v5}$ is an artifact 
of the sampling and frequency.
}
\label{phaserv}
\end{figure}

\section{Stellar Parameters}

$\beta$ Gem is  a nearby K0 III star at a Hipparcos-measured distance of
10.3 pcs (van Leeuwen 2007).  The most recent interferometric measurements for this star 
yielded an angular diameter  of
7.96 $\pm$ 0.09 mas (Nordgren et al. 2001) which corresponds
to a radius of 8.8 $\pm$ 0.1 $R_\odot$.

This star has also been the subject of several spectral investigations. 
McWilliam (1990)
measured an effective temperature, $T_\mathrm{eff}$, of 4850 K, a metalicity of
[Fe/H] $=$ $-$0.07,  and a surface gravity, log $g$ $=$ 2.96.
Gray et al. (2003) derived the same effective temperature, but a lower surface
gravity (log $g$ $=$ 2.52), and higher metalicity ([Fe/H] $=$ $0.08$).
Allende Prieto et al. (2004) determined
$T_\mathrm{eff}$ = 4666 $\pm$ 95 K, a surface gravity of log $g$
$=$ 2.685 $\pm$ 0.09 and a metalicity [Fe/H] $=$ $0.19$.

\begin{table}
\begin{center}
\begin{tabular}{ccccc}
Mode &  $\nu$   & $\nu$ &  Amp &  FAP \\
     &   (c\,d$^{-1}$)  & ($\mu\mathrm{Hz}$) &  (m\,s$^{-1}$)  &  \\
\hline\hline
$f_{v1}$    & 0.8675 $\pm$  0.0001   & 10.03    & 5.39 $\pm$ 0.12   &  $< 10^{-5}$ \\
$f_{v2}$    & 7.5427  $\pm$ 0.0001   & 87.24    & 4.31 $\pm$ 0.10   &  $< 10^{-5}$ \\
$f_{v3}$    & 2.6971   $\pm$ 0.0002   & 31.20   & 2.92 $\pm$ 0.11   &  $< 10^{-5}$ \\
$f_{v4}$    & 8.0404   $\pm$ 0.0001   & 93.02   & 3.99 $\pm$ 0.12   &  $< 10^{-5}$ \\
$f_{v5}$    & 0.6806   $\pm$ 0.0002   &  7.87   & 2.35 $\pm$ 0.14   &  $< 10^{-5}$ \\
$f_{v6}$    & 9.9971   $\pm$ 0.0002   & 115.66  & 2.03 $\pm$ 0.12   &  $< 10^{-5}$ \\
$f_{v7}$    & 4.5330   $\pm$ 0.0002   & 52.41   & 2.48 $\pm$ 0.11   &  $< 10^{-5}$ \\
$f_{v8}$    & 6.9571   $\pm$ 0.0003   & 80.49   & 1.99 $\pm$ 0.11   &  $< 10^{-5}$ \\
$f_{v9}$    & 3.1520   $\pm$ 0.0005   & 36.47   & 1.05 $\pm$ 0.11   &  $< 10^{-5}$ \\
$f_{v10}$ & 11.3124 $\pm$ 0.0003   & 130.88  & 1.50 $\pm$ 0.11   &  $< 10^{-5}$ \\
$f_{v11}$ & 16.8300  $\pm$ 0.0005   & 194.72 & 0.98 $\pm$ 0.11   &  $< 10^{-5}$  \\
$f_{v12}$ & 12.4984 $\pm$ 0.0006   & 144.61  & 0.79 $\pm$ 0.10   &  $< 10^{-5}$ \\
$f_{v13}$ & 6.1863  $\pm$ 0.0005   &  71.58   & 1.05 $\pm$ 0.11   & 0.00001\\
$f_{v14}$ & 1.5158  $\pm$ 0.0006   &  17.54  & 0.90 $\pm$ 0.11   & 0.00016\\
$f_{v15}$ & 17.9305 $\pm$ 0.0006   & 207.46  & 0.87 $\pm$ 0.11   &   0.0074 \\
$f_{v16}$ & 21.7746 $\pm$ 0.0006   & 251.93  & 0.65 $\pm$ 0.10    &   0.0164 \\
$f_{v17}$ & 17.5591 $\pm$ 0.0008   & 203.16  & 0.63 $\pm$ 0.10    &   0.0197 \\
\hline
\end{tabular}
\caption{Pulsation frequencies from the RV data.}
\end{center}
\label{rvmodes}
\end{table}

We also derived the stellar parameters for $\beta$ Gem using a high
signal-to-noise  spectrum taken with the coude echelle spectrograph of the
2m telescope at Tautenburg.
Our analysis yielded  $T_\mathrm{eff}$ = 4835 $\pm$ 50 K, log g = 2.70 $\pm$ 0.10, 
and [Fe/H] = $-$0.07 $\pm$ 0.05. These
values are consistent with the McWilliam and Gray et al. values, and are
in good agreement with the Allende Prieto et al. values with the exception
that our analysis does not indicate such a high metalicity.

The stellar parameters of mass, radius, and age  were determined using
the online tool from Girardi (http://stev.oapd.inaf.it/cgi-bin/param).
This tool uses a
a library of theoretical isochrones (Girardi et al. 2000) and a
modified version of the Baysian estimation method implemented by
Jo{$\!\!\!/$}rgensen \& Lindegren (2005).  A detailed description of this
method is given by da Silva et al. (2006).  As input one needs the
visual magnitude $V$, the parallax $\pi$, the effective temperature
$T_\mathrm{eff}$, and the metalicity [Fe/H]. For the $T_\mathrm{eff}$ and 
the metalicity we used our derived values. The tool yields a 
stellar mass of $M$ = 1.96 $\pm$ 0.19 $M_\odot$,
an age of 1.19 $\pm$ 0.3 Gyr, and a radius of $R$ = 8.29 $\pm$ 0.35
$R_\odot$. The radius is in very good agreement with the one derived from
interferometric measurements. The stellar mass is also consistent with the
frequency of maximum power for the oscillations observed by HZ07.

\begin{figure}
\resizebox{\hsize}{!}{\includegraphics{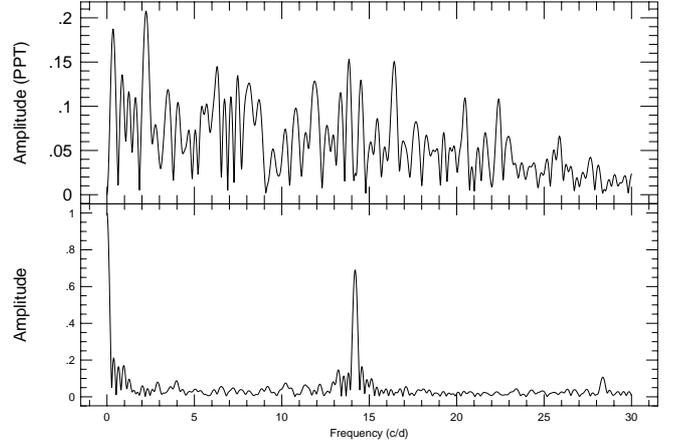}}
\caption{(Top) DFT amplitude spectrum of the MOST photometric data. Units are in parts per thousand (PPT).
(Bottom) the window function. The peak at 14.2 c\,d$^{-1}$ seen in the window function
is due to gaps in the data (see text).
}
\label{mostdft}
\end{figure}

\section{Frequency Analysis}

\subsection{The RV data }

	A frequency analysis was done on the full RV data set 
using the program {\it Period04} (Lenz \& Breger 2004) which performs
a discrete Fourier transform on the data and then fits the time series
with the found frequencies. We are aware that the
oscillations in a giant star may not be coherent over the long time span
of our observations. However, as a first analysis we wanted to see if
a multi-component since function could adequately fit the data,
and more importantly if the derived frequencies had a physical interpretation.
The time unit for our RV analysis program is days, therefore
frequency is expressed in c\,d$^{-1}$. In tables we also give the frequency in
the more common  units of $\mu\mathrm{Hz}$. Figure~\ref{dft} shows
the discrete Fourier transform amplitude spectra for the full data set as
well as the spectral window.
There appears to be excess power in two frequency ranges:
 0.5 -- 3 c\,d$^{-1}$ and 6 -- 10 c\,d$^{-1}$. The highest peak in the
first frequency range is at 0.86 c\,d$^{-1}$ and for the second range at
7.54 c\,d$^{-1}$. The latter corresponds to the dominant frequency found
by HZ07 based on an analysis of a subset of the
data presented here. 

In order to search 
for multiple modes we used the standard pre-whitening procedure. A 
Fourier transform of the data yielded the highest peak in the interval
0 $<$ $\nu$ $<$ 30 c\,d$^{-1}$. A least squares sine fit to the data
was made with {\it Period04} using this 
frequency and its  contribution was subtracted from the data.
A Fourier analysis was then performed on the residuals to find the next
dominant frequency in the data. This process was continued until no additional
significant peaks were found.
Finally, {\it Period04} was used to find a 
solution  by fitting all sine components simultaneously. 

In all cases the highest peak in the amplitude spectra was chosen at each
step of the pre-whitening procedure. We are aware that the 
spectral window, noise, finite lifetime of pulsation modes, etc. can 
cause spectral leakage and the identification of a spurious pulsation mode.
Rather than making a priori judgments as to which peak is real based on a preliminary 
frequency spacing of modes (see below) we simply chose the strongest peak.
This may result in some of our identified modes being an alias frequency
or the presence of artifact frequencies due to finite mode lifetimes.

Our  analysis found a total of 17 significant frequencies in the RV time series. These
are listed in Table 2 (denoted $f_{v1}$ -- $f_{v17}$) in the order in which they were found by the
pre-whitening process. The line in Figure~\ref{timeseries} shows the
fit to the RV data using these 17 modes and Figure~\ref{resids} shows 
the residual RV variations after subtracting the 17-component fit. The final
rms scatter of the data
about this fit is 1.9 m\,s$^{-1}$ for the full data set. The rms scatter
of the data about the final fit for each night is listed in Table 1.
In their analysis of the nights 6--8 HZ07 obtained slightly better fits with
rms scatter of 1.2, 1.5, and 1.7 m\,s$^{-1}$, for each night respectively. The slightly
poorer rms scatter when fitting the entire data set may be due to amplitude variations
in the different modes over the time span of the measurements (116 days) 
or due to mode lifetimes shorter than the full time span of our
observations. If the time scales
of these possible amplitude variations are of $\approx$ days, then fitting subsets
of the data should result in better fits.  The white noise level was estimated
using {\it Period04} in the frequency range $\nu$ = 10--30 $\mu\mathrm{Hz}$
(0.86 -- 2.59 c\,d$^{-1}$) and after
subtracting the contribution of the found modes. This level was $\approx$
0.15  m\,s$^{-1}$.

The statistical significance of the periods were assessed using a
``bootstrap randomization technique'' (see K\"urster et al. 1999).
After removing the previous dominant frequency the RV values were randomly shuffled
keeping the times fixed and a Scargle periodogram calculated
(Scargle 1982). A Scargle periodogram was used since the power is a measure
of the statistical significance of a signal. After a large
number of shuffles (100,000) the fraction of random periodograms
having Scargle power greater than the data periodogram gave an
estimate of the false alarm probability (FAP), that is the probability that 
a signal was due purely to noise possibly in combination with the spectral
window. The FAPs are also listed in Table 2. FAPs with upper
limits indicate that there was no instance in 100,000 random data sets where the
power of the Scargle periodogram exceeded the data periodogram. The first 12 
frequencies found by our pre-whitening procedure are highly significant having 
FAP $<$ 10$^{-5}$. Frequencies $f_{13}$ -- $f_{15}$  are also very significant
having FAPs well below 1\%.
The final two frequencies are moderately significant with FAP $\approx$ 2\%
however, since these fall near the derived frequency spacing (see below)
they may indeed be real.

We compared the amplitude spectrum of the 17-component fit to that of the
observed amplitude spectrum. The top panel in Figure~\ref{final} shows the amplitude spectrum of the
17-component fit to data using the frequencies and amplitudes of Table~2 and
sampled in the same manner as the data. Random noise with $\sigma$ = 2
m\,s$^{-1}$ was also added to the data.
The middle panel of Figure~\ref{final} shows the amplitude spectrum of the final residuals
of the data after removing the 17-component
fit. The bottom panel in the figure shows the amplitude spectrum of
pure noise with a standard deviation of 2.0  m\,s$^{-1}$ that was sampled in the same
manner as the data.

There are several things to note about this figure. 
First, the synthetic fit with
noise has an amplitude spectrum that is similar to  the data spectrum. 
A 17-component sine function with the proper sampling and noise can adequately
reproduce most of the features seen in the amplitude spectrum.
Second, the amplitude spectrum of the final residuals also looks similar to the
spectrum of random noise. Finally, much of the structure seen in the 
data amplitude spectrum at 0.3-3 c\,d$^{-1}$ and 6-10  c\,d$^{-1}$
(top panel Fig.~\ref{dft}) cannot be due to  white
noise combined with the spectral window.

\begin{table}
\begin{center}
\begin{tabular}{ccccc}
HZ07 frequency &   This Work  & Comment \\
($\mu\mathrm{Hz}$) &  ($\mu\mathrm{Hz}$)   &         \\
\hline\hline
29.75  $\pm$ 0.54   & 31.20  & \\
48.41 $\pm$ 1.06    & 36.47  & 1-day alias  \\
79.47  $\pm$ 0.76   & 80.49  & \\
86.91 $\pm$ 0.37    & 87.24  &   \\
104.40 $\pm$ 0.49   & 115.66 & 1-day alias   \\
149.25  $\pm$ 0.63  & 144.61 & 1-day alias + $\Delta\nu_0$   \\
193.63 $\pm$ 3.07   & 194.72 &   \\
\hline
\end{tabular}
\caption{Comparison of frequencies found in this work and HZ07.}
\end{center}
\label{hzcompare}
\end{table}

We now compare the results of our frequency analysis to those of HZ07
whose analysis was based only on a three-night subset of the 
data presented here.
HZ07 found evidence for seven modes in the data and these are listed in Table 3.
This table also lists the nearest frequencies found by the current analysis.
Four of the modes found by an analysis of the full data set
coincide within the errors to peaks found
by HZ07. Two other modes (48.41 $\mu\mathrm{Hz}$ and 36.47  $\mu\mathrm{Hz}$) found by HZ07 are
most likely 1-day aliases of the true frequency. The 149.3  $\mu\mathrm{Hz}$)  mode
found by HZ07 may be an alias of 137.68  $\mu\mathrm{Hz}$. This mode differs by
6.8  $\mu\mathrm{Hz}$ from a mode found by the current analysis. This difference is
close to the frequency spacing, $\Delta\nu$,  found in this work (see 
below). In conclusion,
the frequencies found by HZ07 on the limited subset of data are for the most
part consistent with those found by an analysis of the full data set.

Figure~\ref{phaserv} shows the phase diagram of the RV variations
from the individual modes
found in our frequency analysis. For each displayed frequency the contribution
of the other modes was  first removed before phasing the data. For clarity
we display phase-averaged values using bins of 0.05 in phase. Note that
the phase gap in the panel for $f_{v5}$ is an artifact of the sampling and 
phasing the data to this frequency.

\subsection{MOST Photometry}

A period analysis was also performed on the MOST photometry.
Figure~\ref{mostdft} shows the discrete Fourier transform of the MOST data and the
spectral window.
The white noise in the DFT estimated at frequencies above 30 c\,d$^{-1}$ was about
0.025 parts per thousand (PPT).  The pre-whitening procedure was applied until 
all frequencies with amplitudes
greater than four times the noise level were found. These correspond to false alarm probabilities
less than 0.01.  A total of 16 frequencies were found 
which are listed  in Table 4 (denoted $f_{p1}$ -- $f_{p16}$). 
Also listed are the false alarm probabilities  calculated
using the same procedure as for the RV measurements and 100,000 shuffles of the data.
The final RMS scatter of the data about the fit is 0.13 PPT. 

\begin{table}
\begin{center}
 \setlength{\tabcolsep}{2pt}  
\begin{tabular}{ccccc}
Mode &  $\nu$         & $\nu$     &  Amp &  FAP \\
     &  (c\,d$^{-1}$)  & ($\mu\mathrm{Hz})$ &  (PPT) &                   \\
\hline\hline
$f_{p1}$    & 2.11   $\pm$  0.01   &  24.41 $\pm$ 0.12 & 0.180 $\pm$ 0.013  &  $<$$10^{-5}$   \\
$f_{p2}$    & 0.38   $\pm$  0.01   &   4.40 $\pm$ 0.14 & 0.167 $\pm$ 0.012  &  $\approx$$10^{-4}$ \\
$f_{p3}$    & 6.34   $\pm$  0.01  &  73.35 $\pm$ 0.12 & 0.142 $\pm$ 0.010  &   $\approx$$10^{-5}$   \\
$f_{p4}$    & 7.50   $\pm$  0.01   &  86.77 $\pm$ 0.12 & 0.123 $\pm$ 0.009  &  $\approx$$10^{-5}$ \\
$f_{p5}$    & 8.79   $\pm$  0.01   & 101.70 $\pm$ 0.13 & 0.106 $\pm$ 0.011  &  0.002 \\
$f_{p6}$    & 11.41  $\pm$  0.02  & 132.01 $\pm$ 0.20 & 0.078 $\pm$ 0.017  &  0.002  \\
$f_{p7}$    & 8.18   $\pm$  0.01  &  94.64 $\pm$ 0.15 & 0.120 $\pm$ 0.013  &  0.002 \\
$f_{p8}$    & 3.39   $\pm$  0.01  &  39.22 $\pm$ 0.16 & 0.115 $\pm$ 0.014  &  0.002 \\
$f_{p9}$    & 4.09   $\pm$  0.01  &  47.32 $\pm$ 0.11 & 0.122 $\pm$ 0.010  &  0.002 \\
$f_{p10}$ & 0.88   $\pm$  0.01  &  10.18 $\pm$ 0.14 & 0.138 $\pm$ 0.012  &  0.007 \\
$f_{p11}$ & 2.36   $\pm$  0.03  &  27.30 $\pm$ 0.29 & 0.132 $\pm$ 0.025  &  0.003 \\
$f_{p12}$ & 7.04   $\pm$  0.02  &  81.45 $\pm$ 0.23 & 0.075 $\pm$ 0.020  &   0.002\\
$f_{p13}$ & 1.69   $\pm$  0.02  &  19.55 $\pm$ 0.26 & 0.089 $\pm$ 0.023  &  0.027 \\
$f_{p14}$ & 1.27   $\pm$  0.02  &  14.69 $\pm$ 0.20 & 0.093 $\pm$ 0.017  & 0.010 \\
$f_{p15}$ & 17.84  $\pm$  0.02  & 206.40$\pm$ 0.20 & 0.072 $\pm$ 0.017  & 0.021 \\
$f_{p16}$ & 9.57   $\pm$  0.02 & 110.72 $\pm$ 0.23 & 0.058 $\pm$ 0.020  &   0.010  \\
\hline
\end{tabular}
\caption{Pulsation frequencies derived from MOST photometry. Amplitudes are in parts per
thousand (PPT).}
\end{center}
\label{mostmodes}
\end{table}

\begin{figure}
\resizebox{\hsize}{!}{\includegraphics{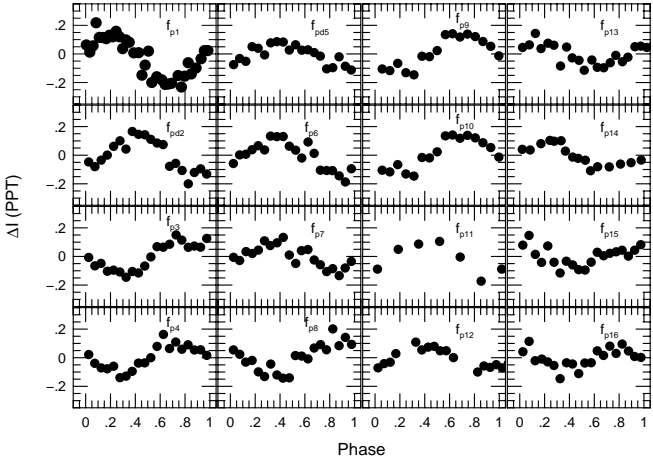}}
\caption{Phase diagrams showing the residual photometric
measurements phased to the
frequencies found in the Fourier analysis. For each panel the contribution
of all other frequencies were subtracted prior to phasing the data. Data were
phased in phase bins of 0.05 except for $f_{p11}$ where 
a bin of 0.1 was used. 
}
\label{photphase}
\end{figure}

\begin{figure*}
\resizebox{\hsize}{!}{\includegraphics{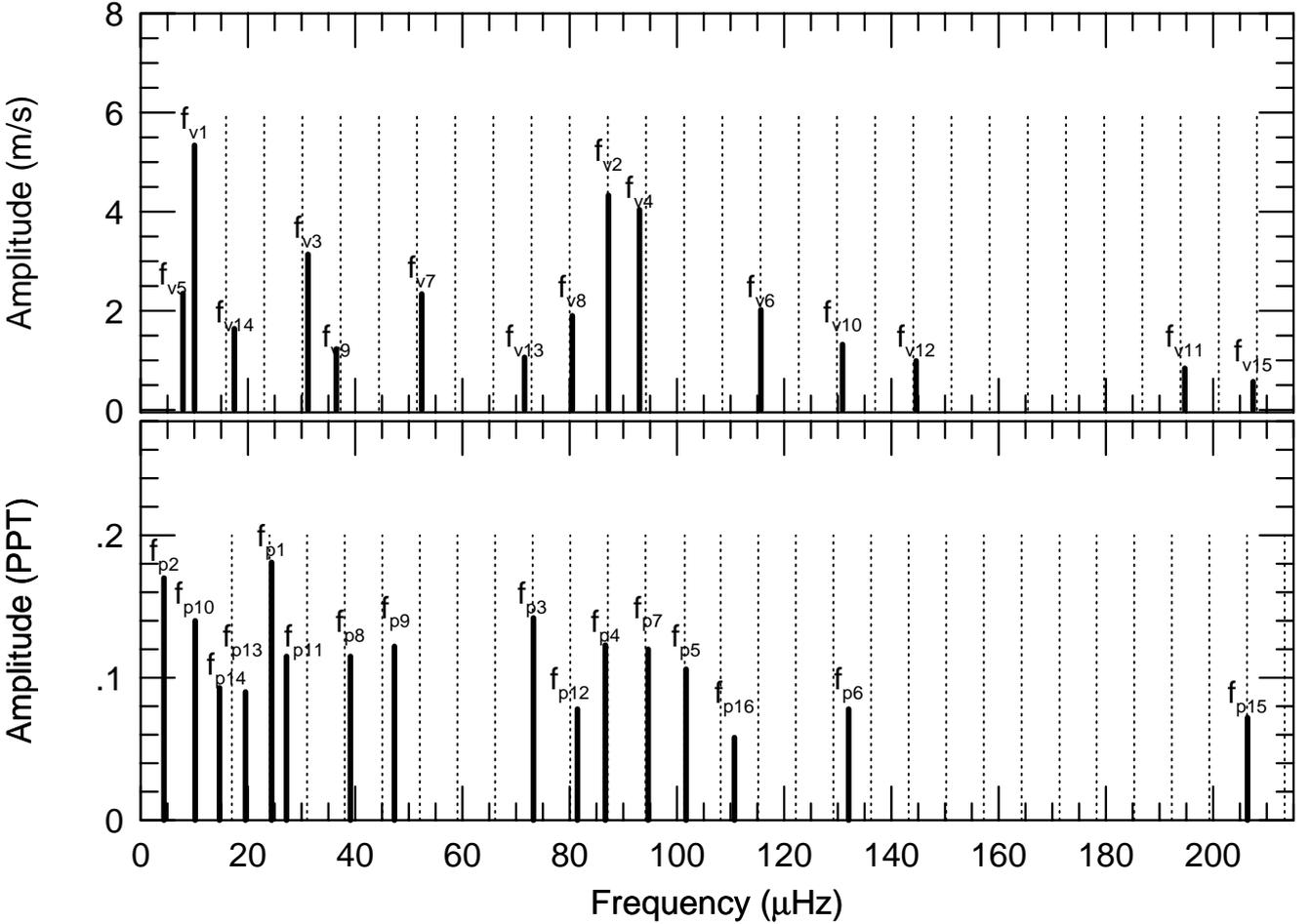}}
\caption{Schematic of the oscillation 
spectrum  for $\beta$ Gem based on the RV measurements (top) and
{\it MOST} photometry (bottom).
}
\label{oscspec}
\end{figure*}

At least six  of the photometric frequencies have counterparts in the RV data.
The highest amplitude photometric modes at 2.11 and 0.38 c\,d$^{-1}$ appear to have no counterparts
in the RV data. However, we caution the reader that a number of effects make a direct comparison 
to the RV frequencies difficult: 1) The RV data are more effected by alias effects than the photometric
data. 2) The spectroscopic and photometric data were  not contemporaneous and the spectroscopic data
in particular were taken over a larger time span. 3) The photometric amplitudes detected by MOST were
small and the time span of the MOST observations were relatively short for the detection of such modes.
A more reliable detection of photometric modes may require a more extended MOST observing run.
In spite of these caveats, it is clear that several of the same oscillation frequencies are seen in
both data sets.

Figure~\ref{photphase} shows the phase diagrams of the individual photometric
modes. As in the case for the RV figure the contribution of all the other
frequencies were first removed before phasing the data. For clarity
we averaged the data in bins of 0.05 in phase. In some cases the
frequency and the sampling resulted in a natural ``clumping'' of points which
dictated a different binning of the data (e.g. $f_{p11}$).

\section{The p-mode Oscillation Spectrum}

The top panel of Figure~\ref{oscspec} shows the schematic of the RV oscillation spectrum (amplitudes and
frequencies) as solid lines. The frequency range 70 -- 100 $\mu\mathrm{Hz}$ seems to 
show modes that are equally-spaced in frequency with a spacing of $\approx$ 10 $\mu\mathrm{Hz}$
However, not all the detected modes seem to follow this spacing. We suspect 
that this may be either 
due to  ``missing modes'' that were not detected either because they have too low
an amplitude, were simply not present, or 
could not be detected due to the sampling of our
data. 

Since we are most likely dealing with
p-mode oscillations it is reasonable to expect that the modes are evenly spaced
in frequency.
We tried to find a mean frequency spacing that could fit the observed spacing.
We took the dominant mode at 87.24 $\mu\mathrm{Hz}$ as our ``origin'' and calculated 
a set of predicted frequencies according to $\nu$ = $\nu_0$ $\pm$ $n\Delta\nu$, where
$\nu_0$ = 87.24 $\mu\mathrm{Hz}$ and $n$ an integer. 
Thus the modes on either side of $\nu_0$ will have $n$ = $\pm$1, $\pm$2, .., etc.
A value 
of $n$ was chosen so that the predicted mode was 
as near as possible in frequency to the
observed modes. A  frequency spacing $\Delta \nu$ was then found that minimized the 
differences between the observed and predicted frequencies.

The dashed lines show the location of the predicted frequencies using 
a frequency spacing of $\Delta \nu$ = 7.14 $\mu\mathrm{Hz}$. Most modes fit extremely
well on this frequency spacing with the possible exception of the lowest 
frequency modes. This is not surprising since only high order modes should have
an evenly-spaced frequency separation. The rms scatter of the observed minus fitted
frequencies (O--C) is 0.38 $\mu\mathrm{Hz}$. Approximately 11 modes were used to derive the
frequency spacing, so we adopt a value of 0.38/$\sqrt{10}$ $\mu\mathrm{Hz}$ = 0.12 $\mu\mathrm{Hz}$ as the error
in the frequency spacing determination.

The bottom panel of Figure~\ref{oscspec} shows the 
schematic of the photometric oscillation spectrum derived from
the MOST photometry. Again we tried to find a mean frequency separation
using the same procedure as the RV data. This resulted in 
a frequency spacing for the photometric modes with 
$\nu$ $>$ 60  $\mu\mathrm{Hz}$ of $\Delta\nu$ = 7.07 $\pm$ 0.32 $\mu\mathrm{Hz}$
consistent with the 
value determined from the RV data.
The grid shows the location of the predicted frequencies found from the RV
measurements.  

\section{The Mass of $\beta$ Gem}

The large separation of p-modes are given by
$$\nu_{n,\ell} \approx \Delta\nu_0 (n + \ell/2 + \delta\nu)$$
where $n$ and $\ell$ are the order and degree of the modes, and
$\delta\nu$ is the small spacing that is related to the stellar structure (Tassoul 1980).
If the modes are non-radial then the observed frequency spacing is just one-half of the
large spacing. If the modes are radial ($\ell$ = 0) then 
the large spacing is the observed frequency spacing.

This large separation is related to the mean density of the star by
$$\Delta\nu = 135 (M/M_\odot)^{0.5} (R/R_\odot)^{-1.5}\hskip 5pt \mu\mathrm{Hz}$$
Assuming that these are all radial modes ($\ell$ = 0), the spacing
of 7.14 $\pm$ 0.12 $\mu\mathrm{Hz}$ and the measured radius of $R$ = 8.8 $\pm$ 0.1 $R_\odot$ results
in a stellar mass of $M$ = 1.91 $\pm$ 0.09 $M_\odot$. 
Using the spacing found in the photometric data results in
$M$ = 1.88 $\pm$ 0.09 $M_\odot$. 
A more precise mass determination requires asteroseismic
modeling of the observed frequency spectrum.

According to the scaling relationships the frequency of maximum power,
$\nu_{max}$, is also related to the mass and radius of the star. From
Kjeldsen \& Bedding (1995):
$$\nu_{max} = M/R^2\sqrt{T_\mathrm{eff}/5777\hskip 2pt K}\hskip 5pt 3.05\hskip 2pt  \mu\mathrm{Hz}$$
where $M$, and $R$ are in solar units. So, in principle one can determine
the mass and radius of the star from $\nu_{max}$ and $\Delta\nu$ alone.
For the case of $\beta$ Gem, $\nu_{max}$ = 0.087 $\mathrm{mHz}$ (ignoring low frequency
modes) and $\Delta\nu$ = 7.14 $\mu\mathrm{Hz}$. This results in $R$ = 9.3 $R_\odot$
and 2.25 $M_\odot$, in good agreement with the interferometric radius and
subsequent mass determination.

We should note that these scaling relations were derived from solar type stars,
but seem to hold also for evolved red giant stars (Kallinger et al. 2009;
Stello et al. 2009).  Also, recent results
from the Kepler Mission indicate that K giants can have mixed $\ell$ = 1 modes (Mosser et al. 2012)
which can shift the observed frequencies. We cannot say from our data whether
$\beta$ Gem has such modes or whether we could have detected them with our RV
measurements. If present they may affect the accuracy of frequency spacing determination.

\section{Discussion}

Our long time series of precise RV measurements for $\beta$~Gem reveals the 
clear presence of p-mode oscillations in this star. These frequencies fall
on a grid of equally-spaced frequency on an interval of $\Delta\nu$ = 7.14 $\pm$ 0.2
$\mu\mathrm{Hz}$. Photometric measurements taken with the MOST micro-satellite also show
a comparable frequency spacing. Most of the photometric  mode match up well with
those detected in the spectroscopy.

It seems that these modes must be high overtone radial modes ($\ell$ = 0).
Nonradial modes would have a large spacing of twice the observed frequency spacing or
$\approx$ 14 $\mu\mathrm{Hz}$. This implies a stellar mass $M$ $\approx$ 8 $M_\odot$ which would be
greatly at odds with stellar evolutionary tracks.  

There is evidence that K giant stars can exhibit   {\it both} radial and nonradial modes.
Hatzes \& Cochran (1994) suggested that the
short-period RV variations in $\alpha$ Boo were consistent with fundamental or low-overtone radial modes.
Buzasi et al. (2000) reported the detection of 10 modes in $\alpha$ UMa using the
guide camera of the failed WIRE satellite. These were spaced on
frequency interval of 2.94 $\mu\mathrm{Hz}$
which was consistent with overtone radial pulsations. Radial modes have also been
claimed for $\epsilon$ Oph (De Ridder et al. 2006) using 
RV  measurements. Barban al. (2007) observed this star 
photometrically with MOST and detected
7 equidistant peaks that were interpreted as radial
modes. However, this result was not without dispute. Kallinger et al. (2008) re-examined the
available data on $\epsilon$ Oph and argued for the presence of 21 modes. They were able
to match these modes with a model that included both radial and non-radial modes. The
unequivocal detection of both radial and non-radial modes finally came from light curves
from the CoRoT space telescope. De Ridder et al. (2009) examined light curves from
giant stars from the first long run of 150 days (LRc01) from CoRoT and
found strong evidence for the
presence of radial and nonradial modes, particularly in the star CoRoT-101034881.
However, Baudin et al. (2012) used 60-d of CoRoT observations of the giant star HD50890 and
found oscillations between 10--20 $\mu\mathrm{Hz}$ that were consistent with a regular spacing of
1.7 $\mu\mathrm{Hz}$ consistent with {\it only} radial modes. So it 
may be that some giant stars can oscillate
in predominantly radial modes and others with both modes. 
(Of course the stars with pure
radial pulsations may also possess non-radial modes that are simply not detected.)

There is some evidence for nonradial modes, at least in the photometric
data. For instance, $f_{p14}$, $f_{p15}$ and possibly $f_{p9}$ may lie
on the half interval of the large spacing derived from what we think are
radial modes.
Nonradial modes may well exist in $\beta$ Gem, but possibly
were  not detected because of their low RV amplitudes and/or
because of our limited time coverage. Furthermore, until the modes
are actually identified some of the ones shown in Figure~\ref{oscspec} may in fact be nonradial
modes.

Evidence for the type of modes (non-radial versus radial) can also come from the
ratio of radial velocity to photometric amplitudes (2K/$\Delta V$). Radial pulsators
such as Cepheids tend to have 2K/$\Delta V$ $\approx$ 50--100 km\,s$^{-1}$ mag$^{-1}$
while nonradial pulsators tend to have
smaller values. Table 5 lists the frequencies in the RV data for which we think we
have found a counterpart in the photometric data. The 2K/$\Delta V$ ratio is listed in the last
column. All but one mode have  2K/$\Delta V$ in the range of 57--103 consistent with radial pulsators.
One mode ($f_{v10}$/$f_{p14}$) shows a 2K/$\Delta V$ ratio consistent
with nonradial modes. We caution the reader that 
because the photometric and spectroscopic data were not simultaneous we cannot be sure that the RV
amplitude for a mode was the same during the time of the photometric measurements. A more
accurate determination of the radial velocity to photometric amplitude would require simultaneous
measurements. However, at face value both the frequency spacing of the modes and the values
2K/$\Delta V$ argue in favor of radial modes.

It is not entirely clear what the nature of the low frequency modes seen in both the
spectroscopic and photometric data is, but low order radial pulsations may  be
a prime candidate. The expected frequencies of the fundamental (F) and first
2 harmonics (1H, 2H) for radial pulsations are 1.5, 2 and 2.5 c\,d$^{-1}$ ($P$ = 0.67,
0.5, and 0.4 d) using the pulsation constants of Cox, King, \& Stellingwerf (1972).
These are close to $f_{v3}$ (2H) and $f_{v14}$  (F) in the RV data and
$f_{p1}$ (2H) and $f_{p11}$ (F) in the MOST data. Using the pulsation constants from
Xiong \& Deng (2007) which may be more appropriate for cool giants results in
a fundamental and first harmonic frequency of 1.67 and 3.22 c\,d$^{-1}$, respectively.
These also close to the frequencies of  $f_{v14}$ and   $f_{v9}$ in the RV data and
$f_{p13}$ and $f_{p8}$ in the photometry.
If these are indeed low order radial modes then it seems that
$\beta$ Gem can excite radial modes over a wide range of orders. We note that the 
low frequency modes $f_{v1}$ and $f_{v5}$ in the RV and $f_{p2}$ and $f_{p7}$
in 
the photometry cannot readily
be attributed to radial pulsations. However, a detailed pulsational analysis
using appropriate stellar models for $\beta$ Gem are needed to identify properly
all detected modes.

\begin{table}[h]
\begin{center}
\begin{tabular}{ccccc}
$\nu_{RV}$ &  Amp$_{RV}$ &  $\nu_{Phot}$   &    Amp$_{Phot}$ & 2K/$\Delta m$ \\
($\mu\mathrm{Hz}$) &  (m\,s$^{-1}$) & ($\mu\mathrm{Hz}$)  &   (milli-mag)  & (km\,s$^{-1}$\,mag$^{-1}$) \\
\hline\hline
80.49 & 2.00 & 81.45 & 0.075    & 53 $\pm$ 14 \\
87.24 & 4.31 & 86.77 & 0.123    & 70  $\pm$ 5\\
93.02 & 4.00 & 94.64 & 0.112    & 71 $\pm$ 6 \\
10.03 & 5.39 & 10.18 & 0.138    & 78 $\pm$ 8 \\
71.58  & 1.05 & 73.35   & 0.140    & 15 $\pm$ 7\\
130.88 & 1.50 & 132.01  & 0.078    & 38 $\pm$ 14 \\
\hline
\end{tabular}
\caption{Ratio of RV to photometric amplitude of modes found in $\beta$ Gem.}
\end{center}
\label{2kamp}
\end{table}

Besides radial versus nonradial modes, the lifetime of the modes in K giant stars has also been the subject of discussion in the literature.
By fitting Lorentz profiles to the power spectrum of 
$\epsilon$ Oph Barban et al. (2007) derived mode lifetimes
of 2.7 days. However, Kallinger et al. (2008) suggested lifetimes of 10 to 20 days and argued that many
of the peaks in the broad Lorentz profiles may have actually been independent modes.
Photometric data on giant stars obtained by CoRoT indeed confirmed that some K giant
stars can have mode lifetimes of tens of days (De Ridder et al. 2009), if not longer. Our excellent fitting of our RV data
for $\beta$ Gem that spans 120 days also suggest the mode lifetimes, at least for this star,
are relatively long, or are re-excited on a regular basis.

The main goal of our analysis was to derive the p-mode frequency spacing of $\beta$ Gem
and to combine this with the interferometric stellar radius in order to determine  a model independent 
stellar mass.  Our resulting stellar mass  of  1.91 $\pm$ 0.09 $M_\odot$
is in excellent agreement  with the value of 1.96 $\pm$ 0.19 $M_\odot$ determined from 
evolutionary tracks.  However, we should note that the mass determination
from evolutionary tracks depends critically on the derived stellar parameters.
For example, if we use the $T_\mathrm{eff}$ and [Fe/H] of Allende Prieto et al. (2004) 
we obtain a stellar mass of $M$ = 1.47 $\pm$ 0.45 $M_\odot$. Thus the actual error in the mass
determination from evolutionary tracks is probably much larger than the formal errors 
using a given set of stellar parameters. The stellar mass derived from the frequency spacing
is thus a better determined value since it is independent of $T_\mathrm{eff}$ and abundance.
$\beta$  Gem is thus one of the few planet hosting stars
for which we have an accurate measurement of the stellar mass.
The combination of stellar oscillations and interferometric 
radius determinations can provide a powerful tool for verifying stellar evolution codes using nearby, bright
giant stars and to obtain more accurate stellar masses for planet hosting giant stars.

The questions regarding whether $\beta$ Gem oscillates only in radial modes, whether
it can also excite nonradial modes,
or what the actual mode lifetimes are all remain open. 
Unfortunately, our RV data set has several long data  gaps and
the MOST data length is too short. Long,
{\it uninterrupted} observations are required for this and this can only be done photometrically from space,
or spectroscopically from multi-site campaigns  or with a dedicated
network of telescopes. This star is an ideal target
for spectroscopic studies with RV measurements with the proposed SONG network of telescopes (Grundahl et al. 2009) or  with
high precision photometry using the micro-satellite BRITE-Constellation (Weiss et al. 2008).

\begin{acknowledgements}
We thank the referee for useful comments that improved the manuscript.
This research has made use of the SIMBAD data base operated
at CDS, Strasbourg, France. APH acknowledges grant HA 3279/5-1 from the
Deutsche Forschungsgemeinschaft (DFG). MZ acknowledges financial support
from DFG grant RE 1664/4-1.

\end{acknowledgements}

\end{document}